\documentclass[reprint,superscriptaddress,amsmath,amssymb,aps,prb]{revtex4-1}
\pdfoutput=1

\usepackage{graphicx}
\usepackage{amsmath}
\usepackage{diagbox}

\begin{document}

\author{Emily V. S. Hofmann}
\affiliation{Department of Electronic and Electrical Engineering, University College London, WC1E 6BT, London, UK}
\affiliation{London Centre for Nanotechnology, University College London, WC1H 0AH, London, UK}
\affiliation{IHP – Leibniz-Institut f\"ur Innovative Mikroelektronik, Im Technologiepark 25, D-15236 Frankfurt (Oder), Germany}

\author{Taylor~J.~Z.~Stock}
\affiliation{London Centre for Nanotechnology, University College London, WC1H 0AH, London, UK}

\author{Oliver~Warschkow}
\affiliation{London Centre for Nanotechnology, University College London, WC1H 0AH, London, UK}

\author{Rebecca~Conybeare}
\affiliation{London Centre for Nanotechnology, University College London, WC1H 0AH, London, UK}
\affiliation{Department of Physics and Astronomy, University College London, WC1E 6BT, London, UK}

\author{Neil~J.~Curson}
\affiliation{Department of Electronic and Electrical Engineering, University College London, WC1E 6BT, London, UK}
\affiliation{London Centre for Nanotechnology, University College London, WC1H 0AH, London, UK}

\author{Steven R. Schofield}
\affiliation{London Centre for Nanotechnology, University College London, WC1H 0AH, London, UK}
\affiliation{Department of Physics and Astronomy, University College London, WC1E 6BT, London, UK}
\email{s.schofield@ucl.ac.uk}


\title{Room temperature donor incorporation for quantum devices: arsine on germanium}

\hyphenpenalty=100000

\date{\today}

\begin{abstract}
Germanium has emerged as an exceptionally promising material for spintronics and quantum information applications, with significant fundamental advantages over silicon. However, efforts to create atomic-scale devices using donor atoms as qubits have largely focussed on phosphorus in silicon. Positioning phosphorus in silicon with atomic-scale precision requires a thermal incorporation anneal, but the low success rate for this step has been shown to be a fundamental limitation prohibiting the scale-up to large-scale devices. Here, we present a comprehensive study of arsine (AsH$_3$) on the germanium (001) surface. We show that, unlike any previously studied dopant precursor on silicon or germanium, arsenic atoms fully incorporate into substitutional surface lattice sites at room temperature. Our results pave the way for the next generation of atomic-scale donor devices combining the superior electronic properties of germanium with the enhanced properties of arsine/germanium chemistry that promises scale-up to large numbers of deterministically-placed qubits.
\end{abstract}

\maketitle

Germanium is experiencing a strong resurgence of interest for the fabrication of (opto)-electronic, spintronic, and quantum technological devices~\cite{Scappucci2020,Hendrickx2021}.  Compared to silicon, germanium has a higher electron mobility~\cite{Lee2005b}, stronger spin-orbit coupling~\cite{Liu1962}, larger Bohr radius~\cite{Liu1962}, larger Stark effect~\cite{Sigillito2016}, and is relatively insensitive to exchange coupling oscillations~\cite{Pica2016,Koiller2002}.  In addition, germanium can be made free of nuclear spin by isotopic enrichment~\cite{Itoh1993}, donors in germanium have long coherence times~\cite{Tyryshkin2012,Sigillito2015}, and germanium is already used in high-performance electronics. 

Electron spins localized on donor atoms in semiconductors form excellent two-level quantum systems (qubits)~\cite{Muhonen2014}.  Assembling individual donors to form devices requires the positioning of potentially thousands of donors with atomic-scale precision~\cite{DeLeon2021}.  This necessitates a detailed understanding of the chemical processes involved, and the ability to produce substitutional donor atoms at atomically-precise locations with extremely high fidelity.  The most successful approach to date is phosphorus in silicon using phosphine (PH$_3$) as the donor precursor, and an atomically-patterned hydrogen resist~\cite{Schofield2003,Schofield2006,Warschkow2016}.  Among the many devices that have been fabricated using this technique are the celebrated single-atom transistor~\cite{Fuechsle2012}, and a few donor device where two-qubit operations were demonstrated via exchange coupling~\cite{He2019}.
 
A critical step in the scale up to large numbers of donor qubits is incorporating the donor atoms into the surface layer~\cite{Schofield2003}. For all of the systems investigated to date---phosphine on silicon~\cite{Schofield2003,Bennett2009} and germanium~\cite{Scappucci2012}, and arsine (AsH$_3$) on silicon~\cite{Stock2020}---a thermal anneal is required to achieve this surface incorporation.  However, this anneal can also provide an opportunity for the donor species to desorb.  The probability of successfully incorporating phosphorus into silicon using a hydrogen resist was studied by two independent groups and determined to be $P_\text{incorp.}\sim70$\%~\cite{Ivie2021,FushsleThesis2011}. Since the probability for successfully fabricating $N$ qubits scales as $(P_\text{incorp.})^N$, this means that even for a modest 50 qubit device, the success rate using phosphorus in silicon can be anticipated to be as low as 1 in $10^{8}$. 

Here, we show that at room temperature, arsenic atoms fully incorporate into the germanium surface layer from adsorbed arsine molecules. Remarkably, and unlike all previously studied systems, no thermal anneal is required for the incorporation of arsenic into germanium.  Thus, the donor incorporation probability is unity, suggesting that arsine on germanium offers a path to the creation of large scale quantum devices consisting of many thousands of donors.  

\section*{Results}
\subsection*{The room temperature dosed surface}

Figure~\ref{fig1}a shows an overview STM image of a Ge(001) surface dosed with arsine at room temperature.  Rows of germanium dimers run horizontally across the image and form either a buckled c$(4\times2)$ configuration or a more symmetric appearing $2\times1$ configuration~\cite{Zandvliet2003}. Features related to arsine dosing are highlighted by boxes. The brightest features that we observe present a single protrusion on top of a dimer row and in the position mid-way between two neighbouring dimers (Fig.~\ref{fig1}b). This feature does not conform to any of the anticipated arsine adsorbate [AsH$_x+(3-x)$H] configurations~\cite{Schofield2006,Warschkow2016,Scappucci2012,Stock2020} (see Supplementary Information), but is an excellent match to a germanium ad-dimer in a ``B'' site configuration~\cite{Zandvliet2000a,Zoethout1998} (Fig.~\ref{fig1}c). Moreover, we occasionally observe these features shifting position on the surface (see white and black arrows on Figs.~\ref{fig1}j--m), which is another known characteristic of germanium ad-dimers at room temperature~\cite{Zandvliet2000a,Zoethout1998,Huijben2006}. The existence of germanium ad-dimers implies that arsenic atoms have incorporated into the surface from the adsorbed arsine molecules, displacing germanium atoms onto the surface in the process. 

\begin{widetext}
\onecolumngrid
\begin{figure}
\centering
\includegraphics[width=\textwidth]{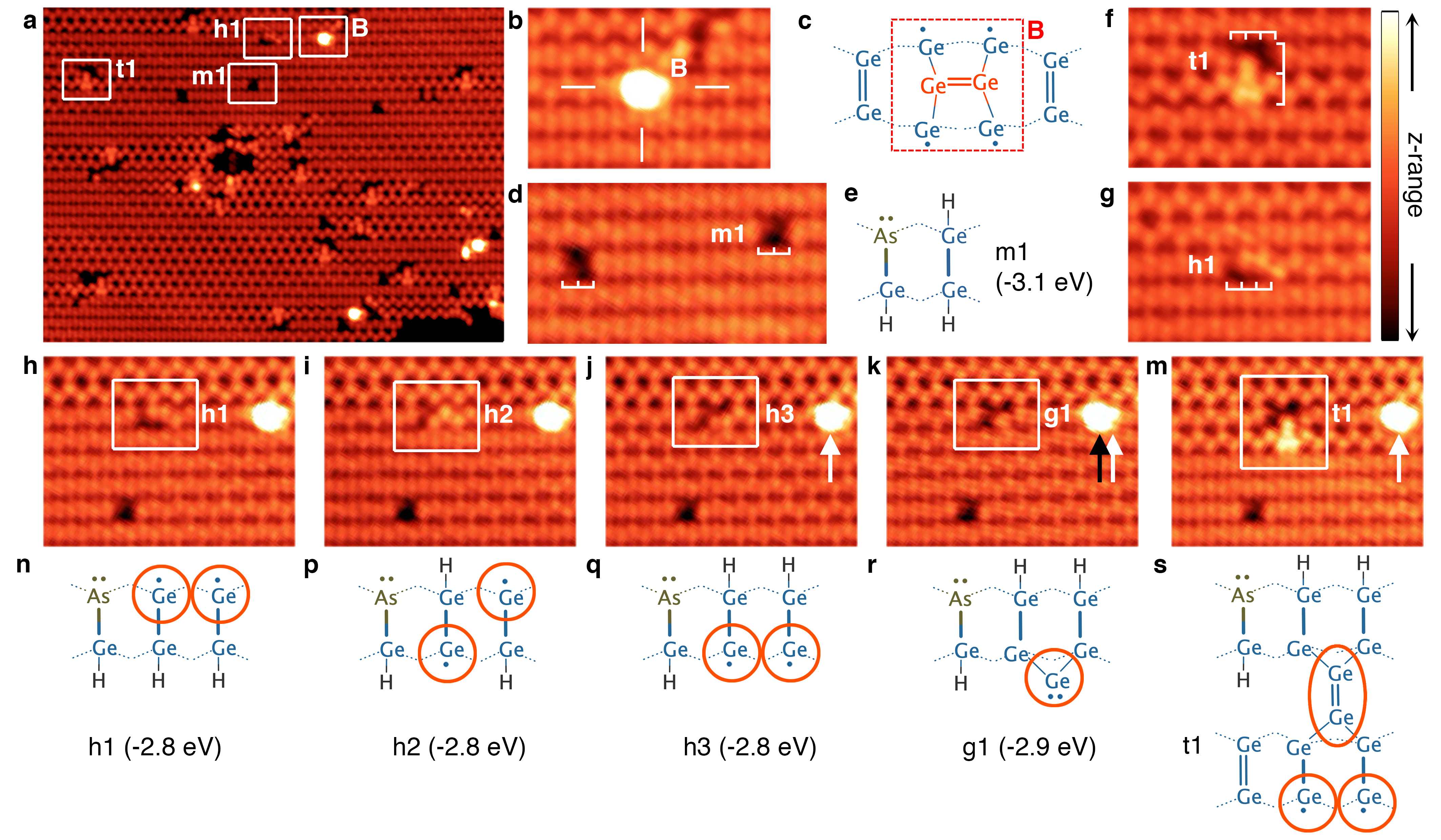}
\caption{\textbf{a,} Overview STM image of arsine (AsH$_3$) on Ge(001) at room temperature. \textbf{b,c,} Germanium B-site ad-dimer image and schematic. \textbf{d,e,} STM image and schematic of an m1 feature. \textbf{f,} Images of t1 feature that is three dimers wide and spans two dimer rows. \textbf{g,} STM image of a h1 feature.  \textbf{h-m,} five images of the same region of the surface where one feature undergoes transitions between h1, h2, h3, g1, and t1 features, and \textbf{n-s,} the corresponding structural schematics.  Calculated DFT energies are quoted for each structure.  Orange circles on panels (\textbf{n-s}) provide a guide-to-the-eye to highlight the location of the brightly-imaging portions of the STM images. STM image parameters: all images $-1.5$~V, 200~pA.  Image (\textbf{a}) z-range 190~pm, images (\textbf{b-m}) z-range 140~pm; the colour scheme is shown top right.  (\textbf{a}) $30\times22$~nm$^2$, (\textbf{b,f,g}) $5.7\times3.8$~nm$^2$, (\textbf{d}) $6.9\times3.8$~nm$^2$,(\textbf{h-m}) $8.0\times6.1$~nm$^2$.} 
\label{fig1}
\end{figure}
\end{widetext}

We next turn our attention to the most abundant feature after arsine dosing: this feature appears as two adjacent dark dimers, and two examples are highlighted in Fig.~\ref{fig1}d.  Close inspection reveals that this feature is not perfectly symmetric, but instead it appears slightly less dark in one corner. We assign this feature to a fully dissociated arsine molecule, resulting in a surface-incorporated arsenic atom and three adsorbed hydrogen atoms forming a hydrogen-terminated arsenic-germanium heterodimer (As--Ge--H) and a germanium monohydride dimer (H--Ge--Ge--H). This structure is illustrated in Fig.~\ref{fig1}e and we label it m1. This feature images less brightly than the surrounding clean germanium dimers due to the absence of the germanium dimer $\pi$-bonds caused by the adsorption of the three hydrogen atoms, and the formation of a lone pair on the arsenic atom.  The slightly less-dark site in one corner of the feature can be attributed to the arsenic atom.

In Figs.~\ref{fig1}f,g we highlight two other arsine features that we label t1 and h1.  Both of these features are three dimers wide. The t1 feature exhibits a complicated appearance and spans two dimer rows (Fig.~\ref{fig1}f), while the h1 feature exhibits a single dark dimer adjacent to two protrusions positioned asymmetrically about the dimer row (Fig.~\ref{fig1}g). We observe transitions between these, and several other three-dimer wide features, as shown in Figs.~\ref{fig1}h--m. By analogy with the m1 feature, we readily assign the h1 feature to a hydrogen-terminated arsenic-germanium heterodimer (As--Ge--H) and two adsorbed hydrogen atoms; however, in this case, the two hydrogen atoms are adsorbed to the ends of two neighbouring dimers to form two germanium hemihydride dimers (Ge--Ge--H), rather than to a single dimer as was the case for m1. The two protrusions observed in the STM image are produced by the two clean germanium atoms of the hemihydride dimers, as highlighted by the orange circles in structural schematic in Fig.~\ref{fig1}n.

In the next image, Fig.~\ref{fig1}i, one of the two asymmetric protrusions has changed from the top of the dimer row to the bottom.  We attribute this change to one hydrogen atom moving from one side of the germanium dimer to the other, as illustrated by structure h2 in Fig.~\ref{fig1}p.  Similarly, in Fig.~\ref{fig1}j we see that the second asymmetric protrusion has now also changed sides, which can be attributed to the second hydrogen moving to the other side of its germanium dimer as shown in structure h3 (Fig.~\ref{fig1}q).

The subsequent transition, observed in Fig.~\ref{fig1}k is, at first consideration, rather unexpected.  Here the two protrusions of the two clean germanium atoms of the hemihydride dimers of structure h3 merge together to form a single protrusion. We explain this change as being caused by the capture of a single germanium monomer at the  clean germanium ends of the two hemihydride dimers in an end-bridge bonding configuration, as shown schematically in Fig.~\ref{fig1}r and labelled g1. 

Figure~\ref{fig1}m shows the final transition in the set, where the feature has now become the two-dimer row spanning feature t1 already mentioned (Fig.~\ref{fig1}f). Here, it appears that the protrusion produced by the monomer of the g1 feature observed in Fig.~\ref{fig1}k has increased in size and intensity, and that our feature has now locally pinned a phason defect (i.e., a defect where two neighbouring dimers become statically buckled in the same direction~\cite{VanHouselt2006,Sweetman2011}) on the adjacent dimer row.  However, we do not expect the spontaneous local pinning of a phason except at very low temperatures~\cite{Sweetman2011,Saedi2009} Thus, we propose a chemical origin for this pinning, specifically the capture of a second germanium monomer to form a dimer-trough bridging germanium dimer, as shown in Fig.~\ref{fig1}s. This structure explains both the enlargement of the protrusion at the site of the monomer in structure g1, and also the dimer pinning that causes the phason-like defect. We further note that the appearance of the trough-bridging dimer of this feature very closely matches that observed in germanium deposition experiments~\cite{Zandvliet2000a,Zoethout1998}.

\begin{figure}
\centering
\includegraphics[width=8.5cm]{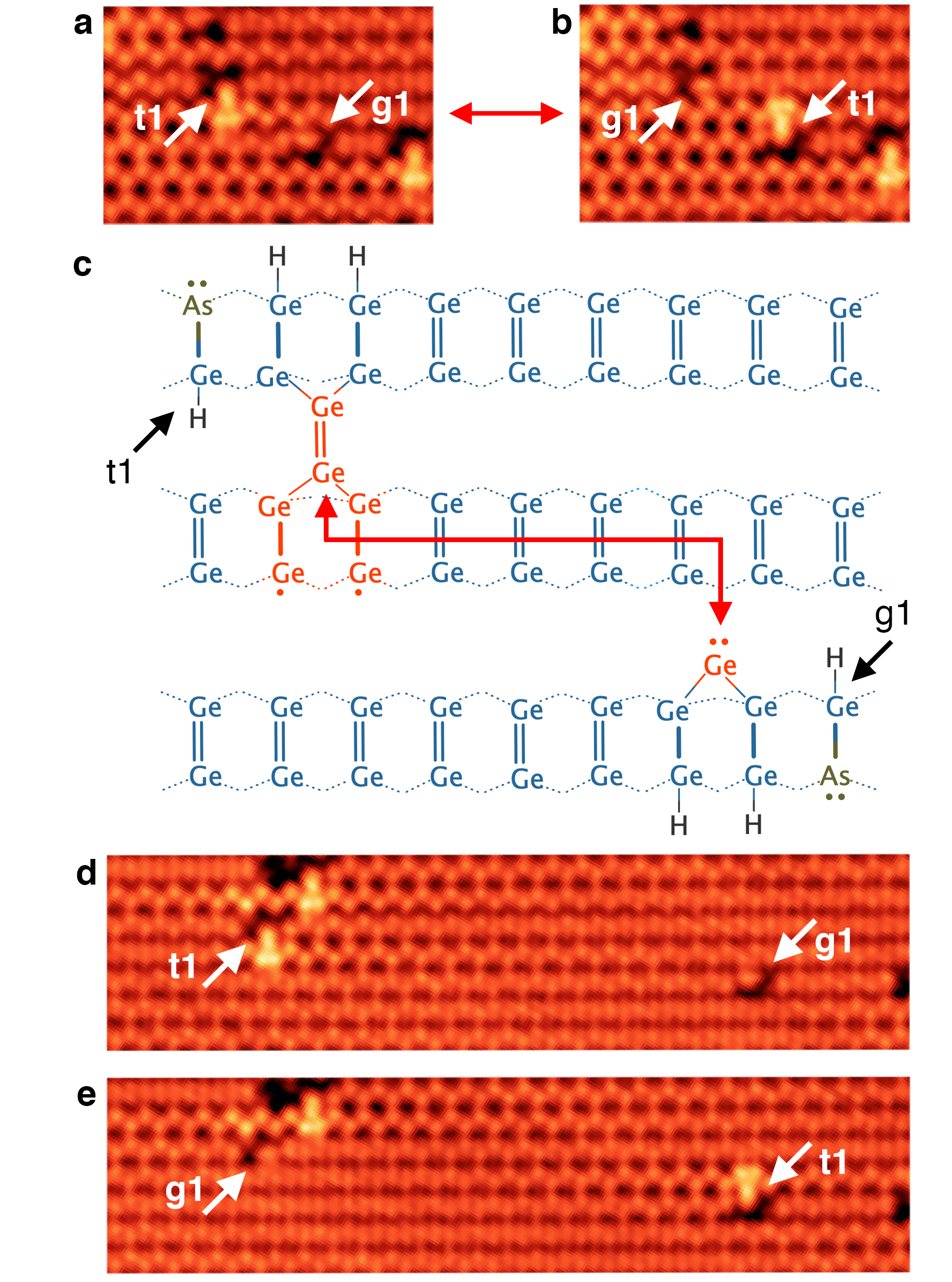}
\caption{Exchange of a germanium monomer between two As-Ge-H heterodimer features (t1 and g1). \textbf{a,} STM image where a t1 feature is seen to the centre-left of the image and a g1 feature is seen centre-right, as indicated.   \textbf{b,} A subsequent STM image of the same region of the surface shown in panel (\textbf{a}); here we see that the t1 feature has changed to g1, and the g1 feature has changed to t1. \textbf{c,} Schematic diagram showing the arrangement and positioning of the t1 and g1 features as they appear in panel (\textbf{a}).  \textbf{d,e,} Two successive STM images where another such monomer exchange was observed between t1 and g1 features.  Image parameters: (\textbf{a,b,d,e}) $-1.5$~V, 200~pA, z-range: 140~pm, (\textbf{a,b}) $8\times5$ nm$^2$, (\textbf{d,e}) $21\times5$ nm$^2$.} 
\label{fig2}
\end{figure}

The transitions $\text{h3}\rightarrow\text{g1}$ and $\text{g1}\rightarrow\text{t1}$ involve the dynamic capture of germanium monomers. These monomers are bound in g1 or t1 features, but are occasionally able to overcome the diffusion barriers at room temperature to leave one feature and be captured at another.  Two examples of such transitions where we identify both the source and the destination of a transiting germanium monomer are shown in Fig.~\ref{fig2}.  Figure~\ref{fig2}a shows an STM image where a t1 and g1 feature are separated by a space of three dimers in the horizontal direction, and a single germanium dimer row in the vertical direction, as illustrated in Fig.~\ref{fig2}c. Figure~\ref{fig2}b shows a subsequent image of the same area of the surface, where we see the t1 feature has become a g1 feature and vice versa, mediated by the exchange of a germanium monomer between them. We observed this exchange several times back and forth in successive STM images, confirming the reversibility of these transitions.  Another example of such an exchange is shown in Figs.~\ref{fig2}d,e.  These observations demonstrate that germanium monomers are highly mobile along the direction of the dimer rows at room temperature in agreement with previous reports~\cite{Zandvliet2000a,Zoethout1998,Swartzentruber1997}; and that monomer capture at a g1 site (to convert it to t1) occurs where the g1 feature exists on the neighbouring row with its monomer directed toward the row on which the mobile monomer is diffusing.  

We gain further insight into the dynamics of the surface by counting the feature transitions between successive STM images.  Table~\ref{table1} shows a count of transitions observed within a $50\times50$ nm$^2$ scan over a period of 12 hours.  There is a strong symmetry to the table about the diagonal, as expected for reversible transitions.  The largest number of transitions, 56\% of the total number of observed transitions (129 out of 231), involve the shifting of hydrogen atoms across the dimer row to cause transitions among h1, h2, and h3 features.  The next largest category are the monomer exchange transitions $\text{g1}\leftrightarrow\text{t1}$, which account for 35\% of the total.  Transitions between h1--h3 and g1 were relatively rare, accounting for 9\% of all transitions; this demonstrates that the majority of monomer transitions are exchanges between dimer t1 and monomer g1 configurations.  No transitions were observed between the h1--h3 and t1 structures, which can be understood since the g1 structure is intermediate between h1--h3 and t1.  Also notable is the absence of transitions $\text{g1}\leftrightarrow\text{g1}$ or $\text{t1}\leftrightarrow\text{t1}$: experimentally these would appear as a mirror symmetry transition about the dimer row, but such transitions are extremely unlikely given our structural assignments.  

\begin{table}[bt]
\caption{Transitions between features in successive STM images over a 12 hour period.  The average time between images was 9 minutes, and the imaging parameters were $-1.5$~V, 200~pA.  We group h1 and h3 into a single feature classification, since experimentally it can be very difficult to delineate between them; transitions between h1 and h3 involve the shifting of both hydrogen atoms of the two hemihydride dimers to the opposite side of the dimer row. Similarly, we note that an alternate configuration of structure h2 exists, where the positions of the two hydrogen atoms are reversed to the opposite side of the dimer row, and we denote this configuration h2$^*$ (see Supplementary Figure S1).}

\label{table1}
\begin{tabular}{|c||c|c|c|c|}
 \hline
 \diagbox{From:}{To:}& \parbox{1.5cm}{h1/h3} & \parbox{1.5cm}{h2/h2$^*$} & \parbox{1.5cm}{g1} & \parbox{1.5cm}{t1}\\
 \hline\hline
 h1/h3 & 15 & 36 & 8 & 0\\
 h2/h2$^*$ & 36 & 42 & 1 & 0\\
 g1 & 11 & 1 & 0 & 40\\
 t1 & 0 & 0 & 41 & 0\\
 \hline
\end{tabular}
\end{table}

We support these compelling experimental observations with DFT calculated formation energies. We calculate each of the experimentally observed structures and find that the formation energies for h1--h3 are all equal at $-2.8$~eV, and structure g1 is nearly equal at $-2.9$~eV.  The closeness of these energies suggests the possibility of reversible transitions between these structures, exactly of the sort that we have already described above.  Structure m1 is slightly more stable at $-3.1$~eV. In addition, we have calculated the energies for a wide range of AsH$_x+(3-x)$H adsorption configurations; the most stable of these are an AsH${}+2$H configuration that we label \textbf{e1} (see Supplementary Figure S1) and an As${}+3$H configuration that we label \textbf{f1}; both have formation energies of $-2.2$~eV.  Thus, we see there is a large energy gain ($0.6$~eV) for the incorporation of arsenic into the top surface layer, and a very large overall energy gain for surface incorporation ($\sim3$~eV) compared to arsine in the gas phase.  These large energy gains are the driving force underpinning the surface incorporation of arsenic into Ge(001) at room temperature.  

As a further confirmation of our interpretation, we check that the densities of ejected germanium and incorporated arsenic are equal.  Feature t1 is one of the most common features that we observe and this accounts for $30\pm11$\% of all arsenic-germanium heterodimer features (m1, h1-h3, g1, and t1) that we observe. The g1 features occur less frequently, accounting for $9\pm4$\% of heterodimer features, while the number of B site germanium ad-dimers is equivalent to $15\pm5$\% of the number of heterodimer features.  Thus, noting that each t1 and ad-dimer feature contains two germanium atoms, we account for $99\pm35$\% of the germanium atoms ejected onto the surface results from  the incorporation of arsenic.

\subsection*{Thermal annealing}

\begin{figure}[t!]
\centering
\includegraphics[width=8.5cm]{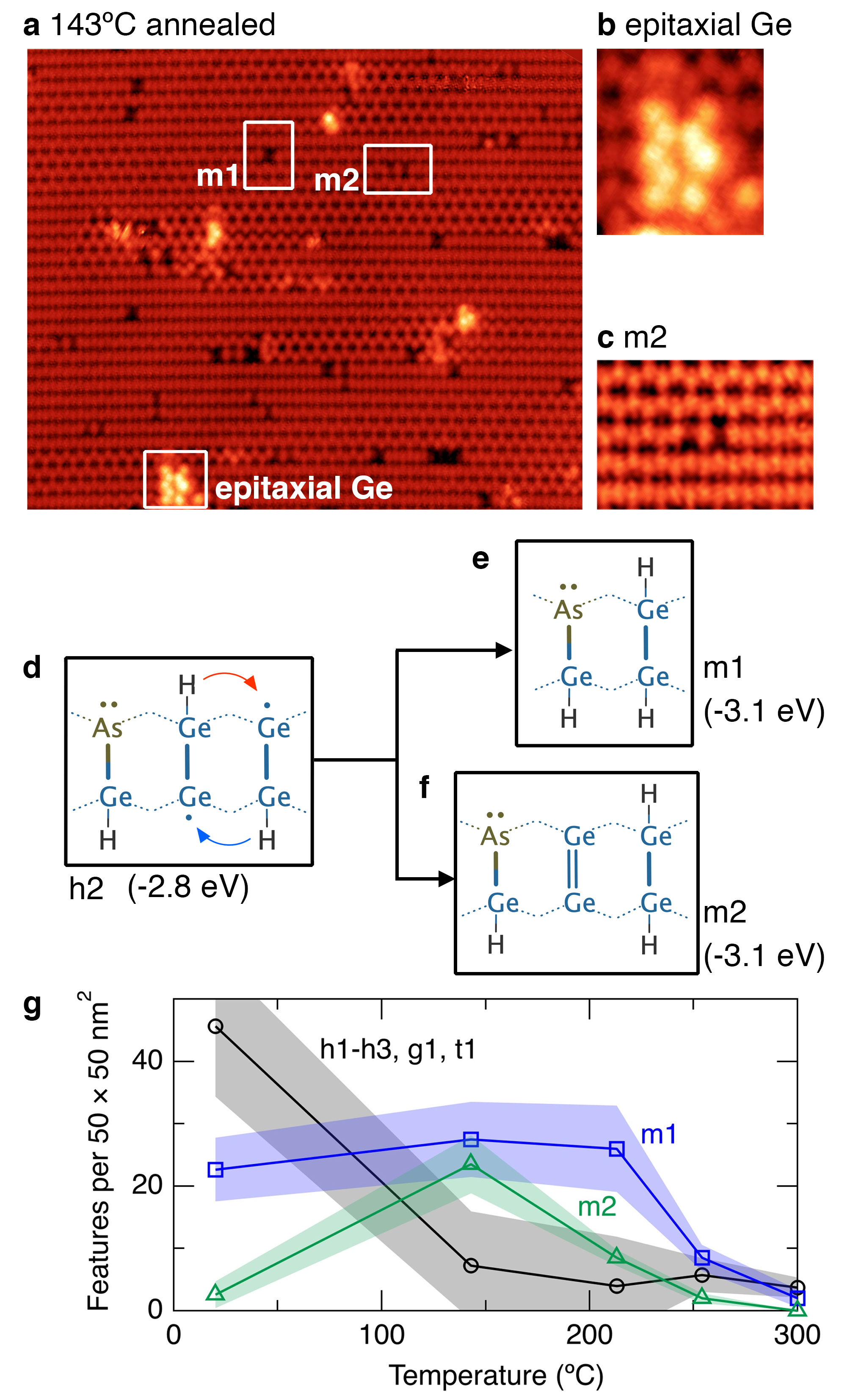}
\caption{\textbf{a,} STM image of arsine dosed Ge(001) after annealing to 140$^\circ$C. \textbf{b,} Epitaxial germanium. \textbf{c,} m2 feature. \textbf{d,} Schematic of a h2 feature highlighting the conversion to \textbf{e,} an m1 feature, and \textbf{f,} a m2 feature.  \textbf{g,} Plot of the feature count of all h1, h2, h3, g1, and t1 features, m1 features, and m2 features as a function of temperature up to 300$^\circ$C. Image parameters: (\textbf{a,b}) $29\times25$ nm$^2$, (\textbf{c}) $6\times4$ nm$^2$, (\textbf{a,b,c}) $-1.5$~V, 200~pA.  z-range: (\textbf{a}) 190 pm (\textbf{b}) 280 pm (\textbf{c}) 110 pm.} 
\label{fig3}
\end{figure}

Figure~\ref{fig3}a shows an image of a dosed surface after annealing to 140$^\circ$C.  Almost all of the room temperature features have disappeared, except the m1 features.  In their place, we find small isolated patches of epitaxial germanium (Fig.~\ref{fig3}b), and features that appear as two dark dimers separated by a single clean germanium dimer (Fig.~\ref{fig3}c).  The presence of epitaxial germanium is evidence that the $140^\circ$C anneal provides sufficient energy for dimer diffusion and the formation of small epitaxial islands, consistent with previous reports~\cite{Li2002,Scappucci2009}.  

The absence of all h1--h3 features demonstrates that this temperature is sufficient for isolated hydrogen atoms in hemihydride (Ge-Ge-H) configurations to shift from one dimer to the next.  If we consider the h2 structure, illustrated in Fig.~\ref{fig3}d, we see two possibilities for such shifts indicated by the red and blue arrows, resulting in the conversion to an m1 feature (Fig.~\ref{fig3}e) or to a new structure, \textbf{m2}, shown schematically in Fig.~\ref{fig3}f. This structure (m2) has a Ge-As-H heterodimer and a single monohydride dimer, making it chemically similar to the m1 feature; however in this case there is a single clean germanium dimer that separates the heterodimer and the monohydride dimer.  The m2 structure has precisely the characteristics of the new features observed in our STM images after 140$^\circ$C annealing (Fig.~\ref{fig3}c), i.e., two dark dimers separated by a clean germanium dimer.

Our DFT calculated energy for the m2 structure is $-3.1$~eV, which is the same as the m1 feature.  Thus, the m1 and m2 features are both 0.3~eV more stable than any of the h1--h3 features; this means the reverse barrier for conversion of m1 and m2 back to h1--h3 is necessarily 0.3~eV larger than the forward barrier to form m1 and m2, explaining the conversion of the h1--h3 structures into m1 and m2. 

In Fig.~\ref{fig3}g we show a plot of the density of features between room temperature and 300$^\circ$C.  We have grouped the features h1--h3, g1, and t1 into a single curve and plot m1 and m2 separately.  The m1 features are the most common feature on the room temperature surface, accounting for $37\pm8$\% of all As-Ge-H features.  After the first anneal (140$^\circ$C) we see the number of h1--h3, g1, and t1 features has fallen almost to zero, while there has been a rise in m1 features and a dramatic rise in m2 features. Upon annealing to higher temperatures up to 300$^\circ$C we see the gradual disappearance of both m1 and m2 from the surface, which can be understood as the process where the hydrogen begins to diffuse away from the m1 and m2 features to form isolated monohydride dimers, leaving hydrogen-free arsenic-germanium heterodimers (As-Ge), as discussed below.

\begin{figure}[t!]
\centering
\includegraphics[width=8.5cm]{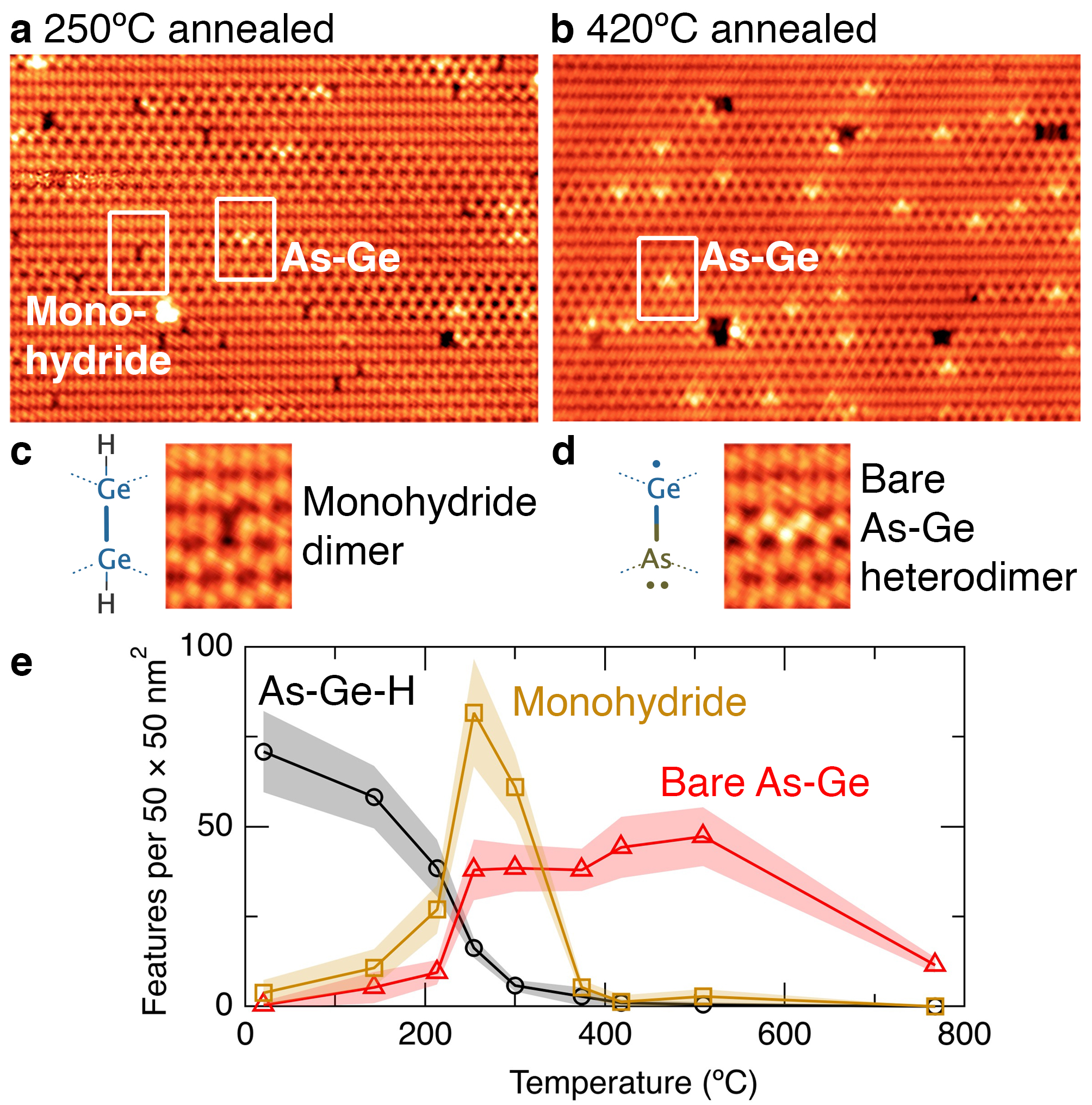}
\caption{STM images of the arsine dosed Ge(001) surface after annealing to \textbf{a,} 250$^\circ$C and \textbf{b,} 420$^\circ$C.  Boxes on panel (\textbf{a}) highlight a monohydride dimer and arsenic-germanium heterodimer, respectively.  \textbf{c,} Structure and STM image of a monohydride dimer.  \textbf{d,} Structure and STM image of a hydrogen-free arsenic-germanium heterodimer.  \textbf{e,} Plot of the feature count of all hydrogen-terminated As-Ge-H heterodimer features (i.e., h1--h3, g1, t1, m1, and m2), mono-hydride features, and hydrogen-free As-Ge heterodimers features up to 800$^\circ$C. Image parameters: (\textbf{a,b}) $27\times19$~nm$^2$, (\textbf{c,d})
$3\times 4$~nm$^2$. (\textbf{a,c,d}) $-1.5$~V, 200~pA. (\textbf{b}) $-1.5$~V, 200~pA. Z-range: (\textbf{a,c,d}) 120~pm, (\textbf{b}) 190~pm } 
\label{fig4}
\end{figure}

Figure~\ref{fig4}a shows an STM image of the surface after annealing to 250$^\circ$C.  In this image we no longer see any of the features from either the room temperature or the 140$^\circ$C annealed surface.  Instead, we see features that appear to be a single dark dimer, highlighted in Fig.~\ref{fig4}c, and features that look like a single dimer that is strongly buckled, as highlighted in Fig.~\ref{fig4}d.  These features are readily identifiable as monohydride dimers, and hydrogen-free arsenic-germanium heterodimers, respectively~\cite{Stock2020,Curson2004}.

Figure~\ref{fig4}b shows an STM image of the surface after annealing to 420$^\circ$C, and here we find the only surviving features are hydrogen-free arsenic-germanium heterodimers.  The absence of hydrogen from this surface is consistent with the known desorption temperature of hydrogen from the monohydride phase on Ge(001) of 330$^\circ$C~\cite{Shimokawa2000}.

We show in Fig.~\ref{fig4}e another plot of the density of features (c.f. Fig.~\ref{fig3}f) on the surface as a function of anneal temperature up to 800$^\circ$C.  Here, we group together all of the features that include a hydrogen-terminated As-Ge-H heterodimer (i.e., h1--h3, m1, m2, g1, t1) into a single curve.  We also plot the density of hydrogen-free arsenic-germanium heterodimers and monohydride.  The density of  As-Ge-H heterodimers diminishes rapidly as we anneal the surface to above $200^\circ$C.  Concomitant with their decrease is a rapid rise in the density of bare arsenic-germanium heterodimers and isolated monohydride.  It is notable that the ratio of hydrogen to arsenic in the range 250--300$^\circ$C is equal to the expected stoichiometric ratio of 3:1 to within our experimental uncertainties (accounting for two hydrogen atoms per monohydride).  The density of bare arsenic-germanium heterodimers remains constant up to at least 500$^\circ$C, but is strongly reduced at our highest anneal temperature (800$^\circ$C), which can be attributed to arsenic diffusion into the bulk~\cite{Stock2020,Solmi1998,Pitters2012}, or arsenic desorption from the surface, or a combination of both.  In the case of desorption, this is likely to be in the form of As$_2$ molecules, as is known to occur for desorption of P$_2$ molecules from the silicon (001) surface between 670 and 820$^\circ$C~\cite{Bennett2010}.

\section*{Discussion}

Our results show arsine on germanium represents a marked departure from all previously studied donor precursors on silicon and germanium. In all previously studied systems an incorporation anneal (typically 350$^\circ$C) is required in order for the donor atoms to adopt substitutional lattice sites in the surface~\cite{Schofield2003,Scappucci2011,Stock2020}.  However, this anneal is highly undesirable as it provides a route for the desorption of the donors from the surface, and presents a major obstacle to the scale up to large-scale quantum technological devices~\cite{Ivie2021,FushsleThesis2011}.  Arsine on germanium solves this problem, since the donor atoms are completely substitutionally incorporated into the surface spontaneously upon adsorption at room temperature, and no incorporation anneal is required.  

Furthermore, we note that the incorporation of arsenic is \textit{rapid}.  We have performed arsine dosing in-situ, allowing us to restart imaging $< 15$ minutes after arsine exposure.  We have not observed evidence for arsenic in any configuration other than As-Ge-H heterodimers. Moreover, in all cases, each As-Ge-H heterodimer is accompanied by two additional hydrogen atoms in the form of either a monohydride dimer, or two hemihydride dimers. This suggests that arsine dissociates and incorporates immediately on adsorption, without any surface diffusion.  Thus, it can be anticipated that arsine will fully dissociate and incorporate into suitably defined lithographic patches with extremely high fidelity, as required for the fabrication of large-scale quantum technological devices, e.g., those involving large numbers of qubits.   

We have presented a comprehensive understanding of the incorporation of arsenic into the germanium (001) surface due to the exposure to arsine at room temperature. We have demonstrated that arsine fully dissociates on the Ge(001) surface at room temperature, and results in the substitutional incorporation of arsenic into the surface layer.  Our results are extremely significant in the context of atomic-scale donor device fabrication as they provide a solution to the donor incorporation probability problem that currently severely limits the scale-up of phosphorus in silicon devices to large numbers of qubits. 

\section*{Methods}

\subsection*{Scanning tunnelling microscopy}

Scanning tunnelling microscopy (STM) experiments were performed in a Scienta Omicron GmbH low-temperature STM system operating at room temperature and under ultrahigh vaccuum $<5\times10^{-10}$~mbar.  High resistivity (1--100~$\Omega$cm antimony doped) germanium 001-oriented samples were degassed overnight at 200$^\circ$C, then heated to $760^\circ$C by direct current heating for 1 hour.  The samples were subsequently cleaned by repeated cycles of sputtering (1.2~kV, 10~mA, 30~min) and direct current annealing (700$^\circ$C, 30~min).  Sample preparation was then completed with three 30~s anneals to $760^\circ$C, with a 25$^\circ$C/min cool down from 600$^\circ$C.  An infrared pyrometer (IMPAC IGA50-LO plus) was used to measure the sample temperature, providing absolute temperature measurements accurate to $\pm30^\circ$C.  Arsine dosing was performed in-situ with the STM tip macroscopically retracted.  The samples were exposed to 99.999\% purity arsine (ATMI Inc.) with a total chamber pressure of $1\times 10^{-10}$~mbar. After arsine dosing, the tip was reapproached and imaging started within 15 minutes of dosing.

\subsection*{Density functional theory}

Density functional theory (DFT) calculations were performed using the B3LYP hybrid exact-exchange functional~\cite{Becke1988,Lee1988}, atom-centred Gaussian-type orbital basis sets, and methods of energy computations and structure optimisation as implemented in the Gaussian 16 software~\cite{Gaussian16}.  A compact Ge$_{21}$H$_{20}$ cluster model was used to represent the Ge(001) surface, describing three Ge-Ge dimers in the surface-closest atomic layer, and eight, four, and three atoms in the second, third, and fourth atomic layer, respectively. The twenty hydrogen atoms of the cluster provide a chemical termination for all germanium atoms other than those of the surface layer. These cluster-terminating hydrogen atoms were held in fixed positions during all geometry optimisations in order to simulate the strain that would be imposed by the surrounding surface and bulk atoms of an extended surface.  All other atoms were fully relaxed during geometry optimisation.

In our calculations two types of composite basis set were used, referred to in the following as \textit{large} and \textit{small}. Geometry optimisations and vibrational frequency calculations were conducted using the small basis, which is composed of the standard 6-311++G(d,p) basis set for all atoms of the adsorbate and the top surface layer of the cluster (i.e. the dimer atoms), a 6-311G(d,p) basis set for second layer atoms, and the core-pseudopotential LANL2DZ basis set for third and fourth layer atoms as well as the cluster-terminating hydrogen atoms. Following geometry optimisation, a single-point energy is calculated at the optimised structure using the large basis set, which is composed of the 6-311++G(2df,2pd) basis set for the adsorbate atoms and top surface layer and a 6-311G(2df,2pd) basis set for all other atoms. Thus, using quantum-chemical notation, the total energy, $E$, of a given structure is calculated as $E_\text{SCF}(\text{B3LYP/large//B3LYP/small}) + \text{ZPE}(\text{B3LYP/small})$.

Adsorption energies, $\Delta E$, of surface structures arising from the adsorption and dissociation of arsine on Ge(001) were calculated as formation energies of adsorption as follows

\begin{equation}
	\Delta E = E_{(\text{cluster+AsH}_3)}-E_\text{cluster}-E_{\text{AsH}_3},
	\label{eq1}
\end{equation}
where $E_\text{cluster}$ is the total energy of the bare germanium cluster, $E_{(\text{cluster+AsH}_3)}$ is the total energy of the cluster with an adsorbed or dissociated arsine molecule, and $E_{\text{AsH}_3}$ is the total energy of a gas phase arsine molecule. 

In some of the adsorption structures considered here, a germanium atom has been displaced away from the adsorption site to a location further away than can be accommodated by our three-dimer cluster. In this case, the adsorption energy is calculated as follows,

\begin{widetext}
\begin{equation}
	\Delta E = \frac{2E_{(\text{cluster+AsH}_3-\text{Ge})}+E_{(\text{cluster}+2\text{Ge})}-3E_\text{cluster}-2E_{\text{AsH}_3}}{2},
	\label{eq2}
\end{equation}
\end{widetext}
where $E_{(\text{cluster+AsH}_3-\text{Ge})}$ is the total energy of the cluster with a dissociated arsine molecule and one missing Ge atom and $E_{\text{cluster}+2\text{Ge}}$ is the total energy of a cluster with a Ge-Ge ad-dimer, bridging perpendicular between two surface dimers. These ad-dimers are directly observed in our experiments and are therefore the correct reference to account for the displaced germanium atoms in our calculations.

\section*{acknowledgement}
This project has been supported by the EPSRC project Atomically Deterministic Doping and Readout For Semiconductor Solotronics (EP/M009564/1). EVSH was partly supported by the EPSRC Centre for Doctoral Training in Advanced Characterisation of Materials (EP/L015277/1), and IHP Leibniz-Institut fu\"r Innovative Mikroelektronik. RC was  supported by the EPSRC and SFI Centre for Doctoral Training in Advanced Characterisation of Materials (EP/L015277/1). TJZS was partly supported by the JSPS/ESPRC core-to-core scheme under project Defect Functionalized Sustainable Energy Materials: From Design to Device Application (EP/R034540/1). SRS acknowledges the use of the UCL Myriad High Performance Computing Facility (Myriad@UCL).

%\section*{Data availability}
%The data created during this research are openly available via zenodo.org at https://doi.org/.

%%%%%%% Bibtex BIBLIOGRAPHY %%%%%%%%
%\bibliography{/Users/steven/academic/tex/bib/library}


\end{document}

\author{Emily V. S. Hofmann}
\affiliation{Department of Electronic and Electrical Engineering, University College London, WC1E 6BT, London, UK}
\affiliation{London Centre for Nanotechnology, University College London, WC1H 0AH, London, UK}
\affiliation{IHP – Leibniz-Institut f\"ur Innovative Mikroelektronik, Im Technologiepark 25, D-15236 Frankfurt (Oder), Germany}

\author{Taylor~J.~Z.~Stock}
\affiliation{London Centre for Nanotechnology, University College London, WC1H 0AH, London, UK}

\author{Oliver~Warschkow}
\affiliation{London Centre for Nanotechnology, University College London, WC1H 0AH, London, UK}

\author{Rebecca~Conybeare}
\affiliation{London Centre for Nanotechnology, University College London, WC1H 0AH, London, UK}
\affiliation{Department of Physics and Astronomy, University College London, WC1E 6BT, London, UK}

\author{Neil~J.~Curson}
\affiliation{Department of Electronic and Electrical Engineering, University College London, WC1E 6BT, London, UK}
\affiliation{London Centre for Nanotechnology, University College London, WC1H 0AH, London, UK}

\author{Steven R. Schofield}
\affiliation{London Centre for Nanotechnology, University College London, WC1H 0AH, London, UK}
\affiliation{Department of Physics and Astronomy, University College London, WC1E 6BT, London, UK}
\email{s.schofield@ucl.ac.uk}


\title{Supplementary Information: \\ Room temperature donor incorporation for quantum devices: arsine on germanium}

\hyphenpenalty=100000

\date{\today}

\begin{abstract}
We present further computational results relevant to the adsorption of arsine (AsH$_3$) on the Ge(001) surface to complement the main text of our manuscript.
\end{abstract}
\maketitle 

\onecolumngrid

\begin{figure}[h!]
\includegraphics[width=\textwidth]{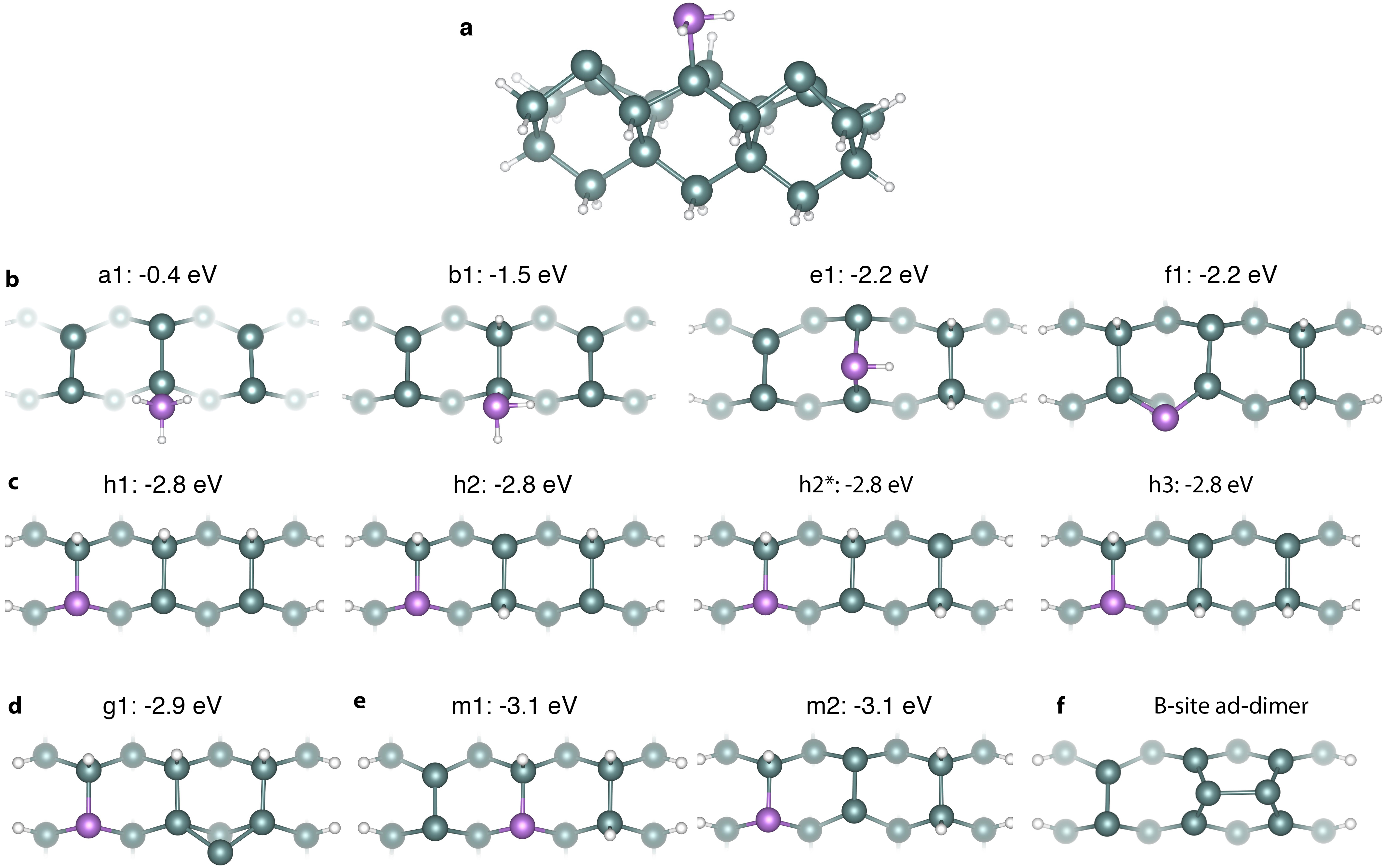}
\caption{\textbf{a,} Perspective view of the $\text{Ge}_{21}\text{H}_{20}$ cluster used in our calculations.  In this illustration an arsine (AsH$_3$) molecule is shown in an $\text{AsH}_2+\text{H}$ dissociative bonding configuration on a single dimer.  \textbf{b,} Top view models of AsH$_3$ in various states of dissociation $\text{AsH}_x+(3-x)\text{H}$. \textbf{c,} Structures h1, h2, h2$^*$, and h3, which all involved surface incorporated arsenic atom forming a hydrogen-terminated As-Ge-H heterodimer and two hemihydride dimers.  \textbf{d,} Structure g1, equivalent to structure h1 with a captured germanium monomer in an end-bridge configuration.  \textbf{e,} Structures m1 and m2 that consist of a hydrogen-terminated As-Ge-H heterodimer and a monohydride dimer.  \textbf{f, } B-site germanium ad-dimer.}
\label{SFig1}
\end{figure}

\noindent Supplementary Figure~\ref{SFig1}a shows a perspective model of the $\text{Ge}_{21}\text{H}_{20}$ cluster used in our calculations.  Details of our computational methodology can be found in the main text.  We have calculated a wide range of structures for the adsorption of arsine (AsH$_3$) on the germanium (001) surface.  Supplementary Figure~\ref{SFig1}b highlights various adsorption structures; adsorption structures a1, b1, e1, and f1 are the most stable structures we find within the classes of AsH$_3$, $\text{AsH}_2+\text{H}$, $\text{AsH}+2\text{H}$, and $\text{As}+3\text{H}$, respectively. Supplementary Figures~\ref{SFig1}c-f show top view schematics for each of the structures discussed in the main text.  

%%%%%%% Bibtex BIBLIOGRAPHY %%%%%%%%
%\bibliography{/Users/steven/academic/tex/bib/library}